# Accessing thermonuclear detonation with the shock front induced by the alpha particle deposition

Bohan Shen[1] [*], Junjue Liao[1] [*], Renjie He[1] [*], Zekun Xu[4,5], Fuyuan Wu[2,3], and Jie Zhang[1,2,3,4] [†]

1  Zhiyuan College, Shanghai Jiao Tong University, Shanghai 200240, China

2 Key Laboratory for Laser Plasmas, Ministry of Education, School of Physics and Astronomy, Shanghai Jiao Tong University, Shanghai 200240, China

3 IFSA Collaborative Innovation Center, DCI joint team, Shanghai Jiao Tong University, Shanghai 200240, China

4 Beijing National Laboratory for Condensed Matter Physics, Institute of Physics, Chinese Academy of Sciences, Beijing 100190, China

5 School of Physical Sciences, University of Chinese Academy of Sciences, Beijing 100049, China

Author to whom correspondence should be addressed: jzhang1@sjtu.edu.cn

## Abstract

The detonation behaviors during thermonuclear burning indicate a state of robust hot spot burning and are widely present in astronomical phenomena, such as supernovae. In this work, we propose an analytical model including alpha-particle deposition at the shock front, which significantly lowers the detonation threshold. The new temperature threshold is 13.4 keV for the isochoric ignition and 25.1 keV for the isobaric ignition, both of which are more accessible experimentally. When a shock wave is present, alpha-particle deposition occurs at the high-density shock front instead of the cold fuel, accelerating the burning wave by approximately 20%. To further validate these findings, we conducted a series of 3D radiation hydrodynamics simulations using finite isochoric hot spots with different fast electron energy. The results reveal a rise in burn-up fraction caused by the detonation wave with a deposited fast electron energy about 8.5 kJ. This work can provide a reference for the realization of fusion energy via fast ignition schemes, such as the double-cone ignition scheme. This work also shows the possibility of studying the detonation in astrophysics with laser driven fast ignition.

## 1. INTRODUCTION

In the hot spot ignition scenario, a small portion of the fusion fuel is first ignited, so that a fusion burning wave can propagate through the cold fuel [1-2]. The isobaric ignition is currently the most widely studied type of ignition. The National Ignition Facility (NIF) has achieved an energy gain > 1 with the isobaric ignition scheme [3-6]. In the isobaric ignition scheme, the pressure in the hot spot formed is equal to the cold fuel pressure, while the hot spot density is significantly lower than the cold fuel [7]. On the other hand, fast ignition schemes have received increasing attention in recent years for its higher energy efficiency [8-12]. Fast ignition separates the compression and heating processes [1,13]. An isochoric high-density plasma is first compressed and then a relativistic electron beam (REB) is injected to form a high-temperature hot spot. The hot spot pressure is significantly higher than that of the fuel [10, 13-15]. For example, the double-cone ignition (DCI) scheme can create isochoric plasmas with a sharp edge in experiments [16]. A fusion gain larger than 100 could be reached via the DCI scheme due to the highly efficient compression and ignition [17].

---

＊ Contributed equally to this work.

The detonation behaviors are very important for an efficient thermonuclear burning propagation and exist widely in astronomical objects such as supernova [18-19]. In the initial stage of burning, there is a characteristic time that signifies the rapid rise in temperature. Since the burning speed is more sensitive to temperature than the sound speed, it increases more rapidly, eventually surpassing the sound speed around this characteristic time. Once the burning speed exceeds the sound speed, detonation may occur under certain conditions. The formation of detonation wave leads to a higher burning propagation speed and burn-up fraction. Although the detonation is very beneficial to the fusion energy, former research has predicted a detonation threshold of over 30 keV by taking into account of the mechanical work of the shock wave alone. This threshold is difficult to achieve in experiments [7]. However, the high-density peak of the shock wave front could stop the α-particles and provides additional energy input, thus lowering the detonation conditions [20]. Till now, the impact of the α-heating in the shock front in thermonuclear detonation remains elusive. And the previous extremely high threshold prevents the investigation of astrophysical detonation waves in laboratory.

In this research, we consider a central hot spot model and analyze the jump conditions at the shock front to calculate the detonation threshold. Regarding the α deposition at the shock front as another energy input, a more accessible value of the temperature threshold is given. Our study also provides a comprehensive description for the thermonuclear burning propagation in the detonation regime.

The paper is organized as follows. In section two, we develop a model of the hot spot evolution during the ignition process. In section three, we analyze the early stage of ignition and give the characteristic time of rapid temperature increasing. In section four, we analyze the jump conditions at the shock front, and give a new criterion for the detonation. The new criterion is verified by O-SUKI-N 3D [21]. In section five, we analyze the different regulation behavior for the burning stage and identified an optimized fast electron energy input for the isochoric model. Finally, a conclusion is given.

## 2. BURNING PROPAGATION MODEL

We consider a central hot spot model as shown in Fig. 1. For simplicity, we assume both hot spot and fuel are uniform during the evolution. The hot spot has time-dependent density $\rho(t)$ and temperature $T(t)$ while the cold fuel has density $\rho$ and temperature $T_0$. The expansion of hot spot $dV$ is divided into two parts: the volume increase $dV'$ caused by the mechanical expansion and the volume increase $dV''=dV-dV'$ of the cold fuel ablated into the hot spot. Here we ignore the interaction between these two parts to reach a uniform state. The corresponding energy equations of the whole hot spot and the cold fuel entering the hot spot are given as follows:

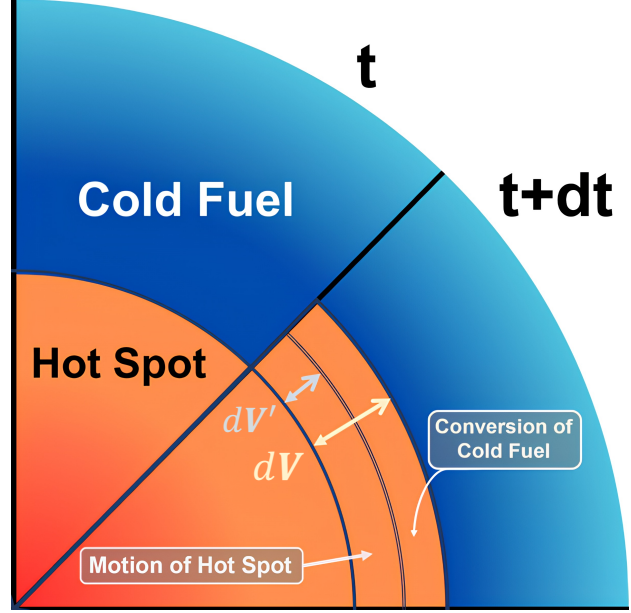

FIG1. Schematic of hot spot formation and propagation

$$\frac{d(eM)}{dt} = W_\alpha - W_r - W_m, \quad (1)$$

$$\frac{d(eM'')}{dt} = W_\alpha(1 - f_\alpha) + W_e, \quad (2)$$

where $e$ is the specific energy, $M$ and $M''$ are the mass of the hot spot and cold fuel entering the hot spot respectively. For the hot spot condition, we have $eM = \frac{3}{2}\Gamma T \cdot \rho V$, where $\Gamma$ is the gas constant and $T$ is the electron temperature. For the DT gas, $\Gamma = 7.66 \times 10^{14} \frac{\text{erg}}{(\text{g·keV})}$. In the first equation, the term $W_\alpha$ is the α-particle heating term, which is given by $W_\alpha = \frac{1}{4\mu^2}\rho^2\langle\sigma v\rangle E_\alpha$. Here $\mu \approx 2.5 m_p$ is the average atomic mass, $\langle\sigma v\rangle$ is the average reactivity and $E_\alpha = 3.5$ MeV is the energy per α-particle. The term $W_r$ corresponds to radiation cooling term [7]. The term $W_m = \Gamma\rho Tu$ is the mechanical work loss and $u$ is the expansion speed. Here we neglect the thermal conduction loss because the thermal conduction power is recycled into the hot spot as the burning wave propagates into the cold fuel. In the second equation, $W_\alpha(1 - f_\alpha)$ and $W_e$ represent the α-deposition and thermal conduction energy into the cold fuel [7]. $W_e = 3A_e T^{\frac{7}{2}}/bR^2 \ln\Lambda$, where $b$ is a numerical coefficient, $R$ is the hot spot radius and the $\ln\Lambda$ is the Coulomb logarithm [22].

To numerically analyze the hot spot evolution, we consider the hydrodynamic effects and give the momentum conservation equation for the fluid as follows:

$$\frac{\partial u}{\partial t} = -\frac{1}{\rho}\nabla P = -\frac{1}{\rho}\frac{P-P_0}{bR}, \qquad (3)$$

where $\nabla P$ is the pressure gradient at the hot spot surface. We use $(P-P_0)/(bR)$ to estimate this gradient. $b$ is the numerical coefficient the same as we use in thermal conduction term. Eq. (3) can be used to solve for the mechanical expansion speed $u$. Adding a kinematical equation:

$$dV' = 4\pi R^2 u dt, \qquad (4)$$

then we can solve the hot spot evolution, as Eq. (1), (2), (3), (4) together form a complete equation set for the hot spot evolution. The numerical results of our proposed model are referred as the numerical solution in the following section.

## 3. EARLY STAGE OF BURNING PROPAGATION

We first discuss the transition process from burning wave to detonation. A semiquantitative analysis of temperature increase at this stage can be obtained from our model. From Eq. (1), (2) we can derive the temperature change ratio as follows:

$$\frac{dT}{dt} = \frac{\left(\frac{1}{4\mu^2}f\langle\sigma v\rangle E_\alpha \rho - A_L \rho T^{\frac{1}{2}} - \frac{3\Gamma T u}{R}\right)}{\frac{3}{2}\Gamma}. \qquad (5)$$

Because the radiation cooling power is about an order of magnitude smaller than the α-heating power when the temperature exceeds 10 keV, the α-particle heating term plays a dominant role in the initial burning propagation process. So, we ignore the radiation cooling term in Eq. (5). The mechanical work term, however, needs some discussion. For the isobaric model, we can have the mechanical expansion speed $u$ as $0$ in short term. While for the isochoric model, a shock wave solution is often used as the mechanical expansion speed. However, if we consider the mechanical expansion in the initial 5 ps, the expansion just begins to accelerate. In this condition, the shock wave solution is no longer suitable. A simple estimation can be made using Eq (3). We consider the typical ignition conditions for the isochoric model, where the $\rho = 300 \text{ g/cm}^3$, $R_h = 20 \text{ μm}$, $T_h = 9 \text{ keV}$, $T_c = 0.5 \text{ keV}$. The acceleration speed then can be estimated as $a = 3*10^{18} \text{ cm/s}$ with $b = 1$. Then we can estimate the expansion speed at $t = 5 \text{ ps}$ as $u = at = 1.5 \times 10^7 \text{ cm/s}$. Based on the former analysis, we further estimate the mechanical work term in Eq. (5) is about $1.6 \times 10^{19} \text{ J/(g·s)}$ and the α-partical heating power term is about $1.7 \times 10^{20} \text{ J/(g·s)}$. So, in a semi-quantitative consideration, we can ignore the mechanical work term in the isochoric model and treat $u$ as $0$. Thus, for both models, we simplify Eq. (5) into the form:

$$\frac{dT}{dt} = \frac{\left(\frac{1}{4\mu^2}f\langle\sigma v\rangle E_\alpha \rho\right)}{\frac{3}{2}\Gamma}. \qquad (6)$$

In the temperature range from 8 keV to 25 keV, the DT reactivity is approximated to be [7]:

$$\langle\sigma v\rangle = \kappa T^2, \qquad \kappa = 1.1 \times 10^{-24} \text{ m}^3/(\text{s}\cdot\text{keV}^2). \qquad (7)$$

In the short term after ignition, the physical quantities of the hot spot exhibit minimal change. To simplify the calculation of temperature variation, the other physical qualities in Eq. (6) can be treated as constants:

$$\frac{dT}{dt} = \frac{\left(\frac{1}{4\mu^2} f E_\alpha \rho \kappa\right)}{\frac{3}{2}\Gamma} T^2, \quad T(t) = \frac{\frac{1}{K}}{\frac{1}{KT_{(t=0)}} - t}, \quad K = \frac{\frac{1}{4\mu^2} f E_\alpha \rho \kappa}{\frac{3}{2}\Gamma}. \tag{8}$$

A rapid temperature increase will occur as the time approaches the characteristic time:

$$t_c = \frac{1}{KT_0} = \frac{\frac{3}{2}\Gamma}{\frac{1}{4\mu^2} E_\alpha \kappa f_0 \rho_0 T_0}. \tag{9}$$

The results agree well with results in ref [2], while we further neglect the change of hot spot density and derive the real time temperature instead of pressure. Note that $t_c$ is inversely proportional to $\rho_0$, indicating that higher hot spot density allows faster temperature increases.

We use Eq. (8) to analyze the isobaric and isochoric ignition under typical parameter conditions and show the significance of hot spot density in rapid temperature increase. The initial ignition temperatures are set the same. The results are shown in Fig. 2(a). The trends of numerical curve and Eq. (8) fits well, revealing that the Eq. (8) provides a decent estimation of the temperature rise. We also plot the corresponding characteristic time in the Fig. 2(a). The characteristic time for the isochoric ignition is about 5.5 ps while for the isobaric ignition it is about 19 ps, about 3 times of the isochoric $t_c$.

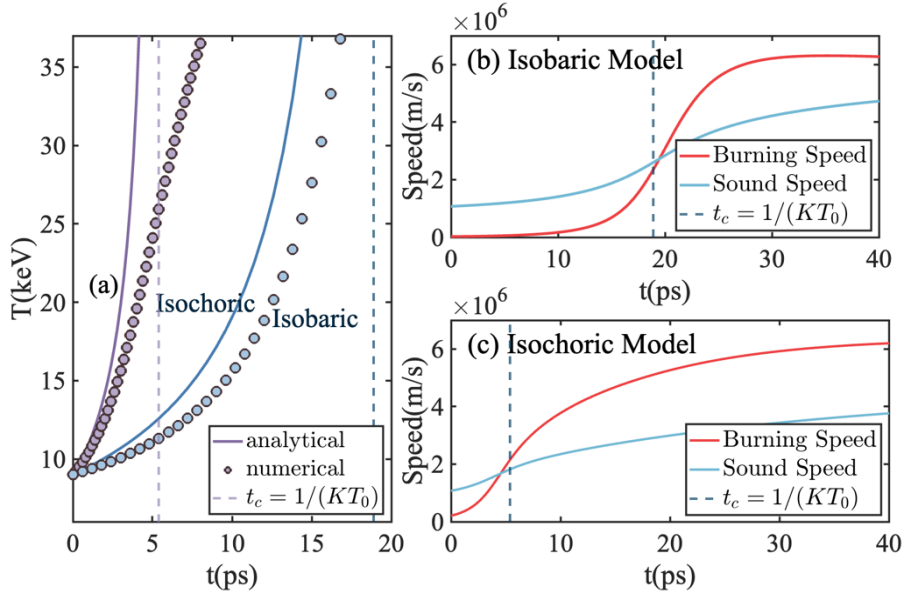

FIG 2. For the isobaric model, T=9 keV, ρ=100 g/cm³, R=30 μm. For the isochoric model, T=9 keV, ρ=300 g/cm³, R=20 μm. (a) The numerical result of T-t (blue line), analytical solution without bremsstrahlung (circle icon) and 1/(KT$_0$) (vertical dotted line) for the isochoric and isobaric model. (b)(c) The comparisons of sound speed (blue line), burning wave speed (red line) and 1/(KT$_0$)

We further explore the competition between burning wave and rarefaction wave in the early stages of burning propagation. The rarefaction wave propagates with sound speed. The comparison of two speeds in ignition process are also shown in Fig 2(b)(c). Both the isochoric and isobaric ignition starts with burning speed slower than sound speed, but the burning speed in an isochoric ignition exceeds the sound speed about three times faster than that in an isobaric ignition. We also plot the $t_c$ in the figures. Note that the time burning speed exceeds sound speed is close to the characteristic time. This can be explained simply as follows: The burning speeds of both models are proportional to the DT reactivity. In temperature range 8-25 keV, the burning speed is proportional to $T^2$, while the sound speed is proportional to $T^{\frac{1}{2}}$. So, as the time approaches the characteristic time, the temperature starts to increase rapidly, so as the burning speed and the sound speed. As a result, $t_c$ could also be regarded as a mark of the exceeding. It should be noted that the breakdown of subsonic assumption implies the existence of a detonation, which will be further discussed in the next section.

## 4. Detonation Regime

During the burning process, the higher hot spot pressure drives a shock wave into the surrounding cold fuel. If the temperature reaches a sufficiently high level, the detonation will occur. The detonation is defined as follows: as the shock wave passes through a portion of cold fuel, the mechanical work heats the fuel to ignition temperature. Thus, the front of the shock wave can be also regarded as the front of the burning wave. An approximate detonation condition is given by [7]:

$$T_h > \frac{4\rho_c}{\rho_h} T_{ign} \approx 30 \frac{\rho_c}{\rho_h} keV. \qquad (10)$$

However, the definition above and Eq. (10) ignore the α-deposition. The high-density shock front stops the α-particles and deposit their energy. The α-deposition provides an additional heating to the fuel, making the detonation easier to access.

We simplify our burning propagation model to a one-dimensional shock model as shown in Fig. 3. We assume the hot spot described by the density $\rho_h$, temperature $T_h$ and pressure $p_h$. The cold fuel characterized by a density $\rho_0$, temperature $T_0$ and pressure $p_0$. When the shock wave reaches the cold fuel, the cold fuel turns into a state with density $\rho_1$, temperature $T_1$ and pressure $p_1$ momentarily.

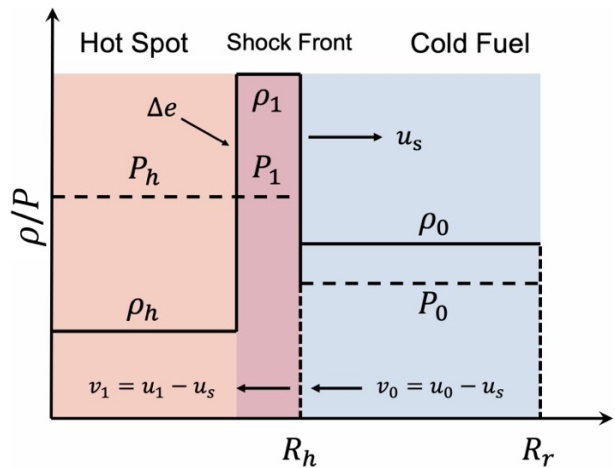

FIG 3. Schematic of one-dimensional shock model. From left to right are hot spot area, shock front area and cold fuel area. $R_h$ represents the radius of hot spot and $R_r$ represents the radius of the whole fuel.

Assuming the pressure in the hot spot is comparable to the pressure of shock front [7], we can write $p_h = \rho_h \Gamma T_h = p_1 = \rho_1 \Gamma T_1$. With an additional power input from α-particle deposition denoted as $\Delta e$, we can write the energy conservation equation as follows:

$$\frac{p_1 V_1}{\gamma - 1} - \frac{p_0 V_0}{\gamma - 1} = \frac{1}{2}(p_0 + p_1)(V_0 - V_1) + \Delta e, \tag{11}$$

where $\gamma = 5/3$ is the adiabatic exponent and $V_0$ and $V_1$ are the specific volumes written by $1/\rho_0$ and $1/\rho_1$ respectively. A simplification of the equation gives:

$$\frac{\rho_1}{\rho_0} = \frac{(\gamma + 1) + (\gamma - 1)\frac{p_0}{p_1}}{(\gamma - 1) + (\gamma + 1)\frac{p_0}{p_1}} - \frac{2(\gamma - 1)\Delta e}{\left[(\gamma + 1)\frac{p_0}{p_1} + (\gamma - 1)\right] p_1 V_1}. \tag{12}$$

As the hot spot pressure is much higher than the fuel pressure, we use the strong shock approximation $p_1/p_0 \gg 1$, which gives:

$$\frac{\rho_1}{\rho_0} = \frac{(\gamma + 1)}{(\gamma - 1)} - \frac{2\Delta e}{p_1 V_1} = \frac{(\gamma + 1)}{(\gamma - 1)} - \frac{2\Delta e}{\Gamma T_1}. \tag{13}$$

The above equation gives the density after the shock as $\rho_1 = \rho_0 \left(4 - \frac{2\Delta e}{\Gamma T_1}\right)$. Using the assumption $\rho_h T_h = \rho_1 T_1$, we can obtain the temperature after the shock as:

$$T_1 = \frac{\rho_h T_h}{4\rho_0} + \frac{\Delta e}{2\Gamma}. \tag{14}$$

To achieve the detonation, the fuel temperature must exceed the ignition temperature, which gives $T_1 > T_{ign}$. Then we have:

$$T_h > \frac{4\rho_0}{\rho_h}\left(T_{ign} - \frac{\Delta e}{2\Gamma}\right). \tag{15}$$

It is shown in Eq. (15) that the additional α-particle deposition energy can lower the hotspot temperature needed for the detonation. Then we derive the additional energy input $\Delta e$ by the α-deposition. For simplicity, assuming that the α-particle deposition is instantaneous, and the α-particles that originally escape the hot spot all be stopped by the shock wave front, then the energy input power is given by:

$$P_\alpha = \frac{1}{4\mu^2}(1-f)\rho_h^2 \langle \sigma v \rangle E_\alpha V. \tag{16}$$

The additional energy input satisfies $P_\alpha = \rho_0 v_0 S \Delta e$, where $v_0$ stands for the speed of shock wave front. Thus, the energy input $\Delta e$ can be expressed as follows:

$$\Delta e = \frac{\frac{1}{12\mu^2}(1-f)\rho_h R \langle \sigma v \rangle E_\alpha}{v_0} \frac{\rho_h}{\rho_0}. \tag{17}$$

From Eq. (13) along with the conservation of mass and momentum in the one-dimensional shock model, we can deduce the shock wave speed in the strong shock

approximation as:

$$v_0 = \sqrt{\frac{4-\beta}{3-\beta}\Gamma T_h \frac{\rho_h}{\rho_0}}, \quad \beta = \frac{2\Delta e}{\Gamma T_1}. \tag{18}$$

Eq. (14), Eq. (17) and Eq. (18) together gives the equation of $\Delta e$:

$$A\Delta e^2 + B\Delta e + C = 0, \tag{19}$$

where $A = \left(\frac{\Gamma\rho_h T_h}{\rho_0}\right)^2$, $B = \frac{1}{2}\left(\frac{1}{12\mu^2}(1-f)\rho_h R\langle\sigma v\rangle E_\alpha \frac{\rho_h}{\rho_0}\right)^2$ and $C = -\frac{3\Gamma\rho_h T_h}{4\rho_0}\left(\frac{1}{12\mu^2}(1-f)\rho_h R\langle\sigma v\rangle E_\alpha \frac{\rho_h}{\rho_0}\right)^2$.

The threshold of the detonation for the isobaric and isochoric model can be obtained based on the typical density ratio value, shown in Fig. 4(a). For a hot spot and cold fuel with given temperature $T_h$ and $T_0$, given density $\rho_h$ and $\rho_0$, the additional energy input $\Delta e$ can be obtained from Eq. (19), then the threshold can be gained by $\Delta e$ from Eq. (15).

Compared to the solution of shock wave speed $v_0'$ obtained by neglecting the α-deposition energy, $v_0$ is consistently higher. While for the isobaric model, the difficulty of achieving detonation still exists, the detonation state is accessible for the isochoric model with a temperature of 13.4 keV [Fig. 4(a)]. For the typical parameters of the isobaric model achieving detonation state (T=13.4 keV, ρ=300 g/cm³, R=20 μm), the ratio $\frac{v_0}{v_0'} = \sqrt{\frac{12-3\beta}{12-4\beta}} = 1.21$. So, at the early stage of burning propagation, the α-deposition can accelerate the process of the isochoric model by approximately 20%, advancing a more rapid propagation of the thermonuclear burning.

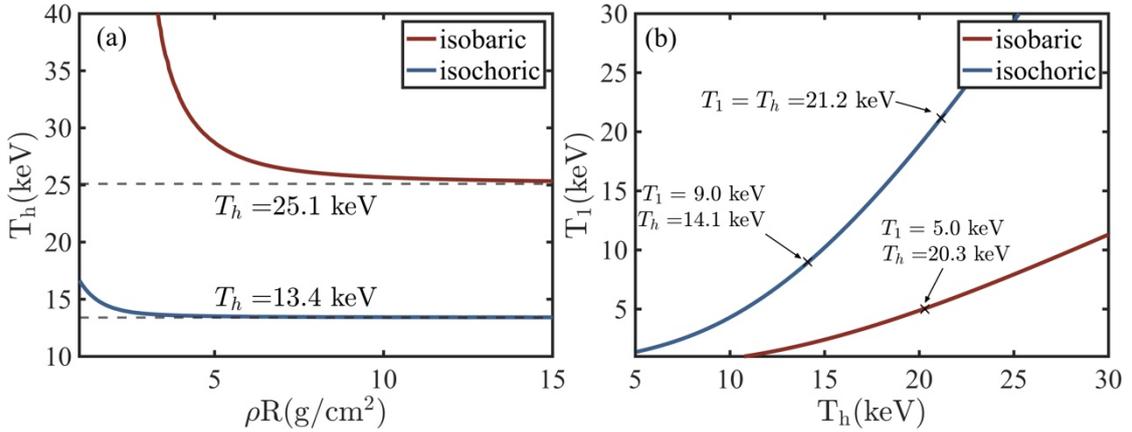

FIG. 4. (a) The threshold for forming a detonation as a function of hot spot temperature and hot spot area density. ($T_{ign} = 7\ keV$ to compare with Atzeni's results) (b) Relation of shocked fuel temperature $T_1$ and hot spot temperature $T_h$ under simplification of Eq. (18) is plotted.

From Fig 4(a), threshold temperature for the hot spot to achieve the detonation approaches a certain value as the areal density increases. If we use the ignition condition $T_{ign} > 7$ keV as in ref 7, the detonation condition is $T > 13.4$ keV for the isochoric model and $T > 25.1$ keV for the isobaric model, both considerably lower than previous predictions $T > 30$ keV. The asymptotic behavior can be explained as follows. After some time of fusion reactions, the α-particle range is close to the hot spot radius because of the self-regulating effect. Then the normalized hot spot radius $\tau_\alpha \approx 1$, the deposition ratio $f$ thus can be approximate to [23]:

$$f = 1 - \zeta \frac{T_h^{\frac{3}{2}}}{\rho_h R}, \quad \zeta = \frac{0.107}{4 \ln \Lambda}, \tag{20}$$

The fraction of α-particle that deposited in the shock wave front thus can be expressed as $1 - f = \zeta T_h^{\frac{3}{2}}/\rho_h R_h$. So, the extra energy input described by equation (17) can be rewritten as follows:

$$\Delta e = \frac{\frac{1}{12\mu^2} \zeta T^{\frac{3}{2}} \langle \sigma v \rangle E_\alpha}{v_0} \frac{\rho_h}{\rho_0}, \tag{21}$$

which is uncorrelated to areal density. With the simplified energy input Eq. (21), we calculate the relation between temperature of the shock front and the hot spot temperature for both models, shown in Fig. 4(b). Two specific temperatures for the isochoric model are marked in the figure. The first represents the detonation condition, where $T_1$ reaches the ignition temperature 9 keV for the isochoric model and 5 keV for the isobaric model. The second represents the point where the cold fuel is directly heated to the hot spot temperature by the shock.

The physical meanings of these two points are as follows. After the ignition, the fusion reaction drives a shock outward. The shock speed is faster than the burning speed. At the first point, the thermonuclear burning propagation transfers into a detonation, intensifying the fusion reaction. Different from previous perspective, we regard this point as the symbol that the burning wave begins to catch up, as the catch-up process will not be instantaneous practically. At the second point, the shock directly heats the fuel to the hot spot temperature and makes it part of the hot spot. So, we regard this point as the symbol that the burning wave finally catch up with the shock wave, and after which the burning wave may even exceed the shock wave. In Fig. 4 (b), the burning wave catches up at hot spot temperature 21.2 keV for the isochoric model, while it never catches up for the isobaric model.

We use 3D radiation hydrodynamics code O-SUKI-N to simulate the burning propagation process. The temperature and density distribution for two models at different times are shown in Fig. 5. For both models, a density peak, regarded as the shock wave front, is driven into the cold fuel after the ignition. We define the burning wave front as where the temperature exceeds 7 keV and the shock front as where the density is the highest.

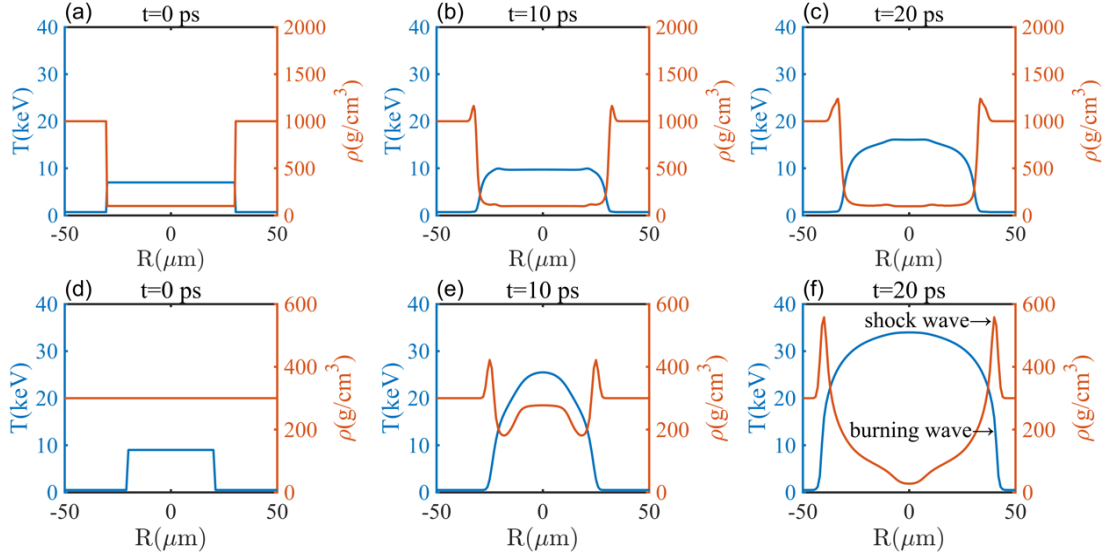

FIG 5. (a) (b) (c) The temperature and density distribution at 0 ps, 10 ps, 20 ps, respectively for the isobaric model. The initial conditions of hotspot are T=7 keV, ρ=100 g/cm³, R=30 μm. (d) (e) (f) The temperature and density distribution at 0 ps, 10 ps, 20 ps, respectively for the isochoric model. The initial conditions of hotspot are T=9 keV, ρ=300 g/cm³, R=20 μm.

The simulation results agree well with our theoretical analysis, as shown in Fig. 6. For the isobaric model, the burning wave trends to chase the shock wave, but never catches up in 35 ps. While for the isochoric model, the figure clearly shows the existence of the detonation in the early stage of burning propagation. The burning wave starts to chase after the shock wave around 5 ps and finally catch up at 13.7 ps. Notably, the mean hot spot temperature at catch-up moment is 21.2 keV, the same with our theoretical prediction.

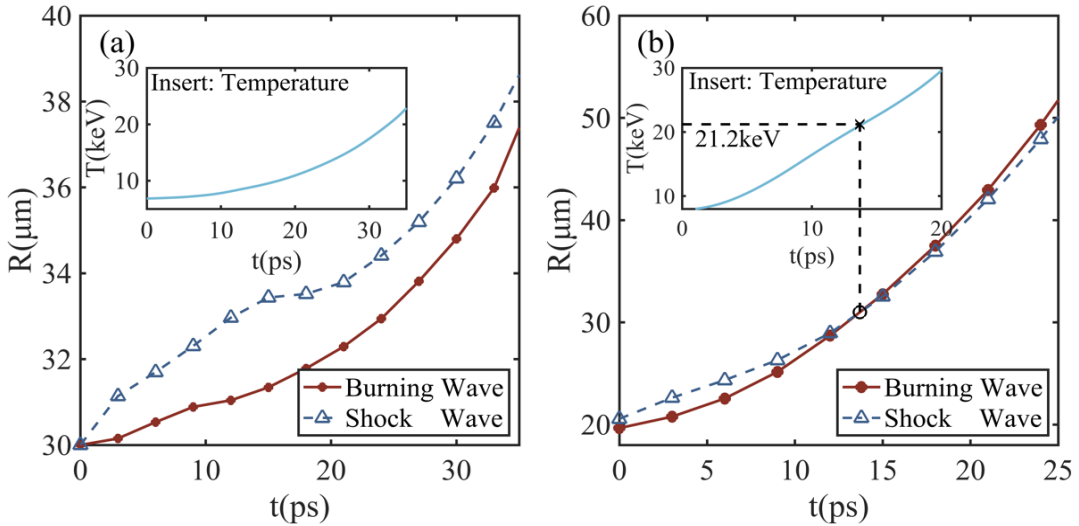

FIG 6. Burning wave radius (red line with circle icon) and shock wave radius' (blue line with triangle icon) relations with time for the isobaric model (a) and the isochoric model (b). And the isobaric (isochoric) initial conditions are T=7 keV, ρ=100 g/cm³, R=30 μm (T=9 keV, ρ=300 g/cm³, R=20 μm) Inset: the average hot spot temperature of the isochoric model and the isobaric model.

# 5. Analysis of Self-Regulation Behavior and Deposited Fast Electron Energy

The transition from normal burning wave to the detonation is affected by the self-regulation behavior. The range of α-particles can be expressed as $\rho l_\alpha = \frac{0.107\, T_e^{1.5}}{\ln \Lambda}$, and the normalized hot spot radius is defined as $\tau_\alpha = \frac{R_h}{l_\alpha}$ [7,23].

In the early stage of the isobaric model, high hot spot temperature results in larger $\alpha$-range and lower $\alpha$ particles deposition fraction ($f$) in the hotspot, which is beneficial for the formation of detonation, as shown in Eq. (17). However, the self-regulation behavior would suppress the temperature rise, which can be seen from Fig 7(a). The high $\rho R$ cases have the similar negative feedback. The negative feedback increases the threshold of detonation and prevents the burning wave to catch up, which is shown in Fig. 6 (a), where the gap between the burning wave and shock wave initially increases, then decreases, and finally stabilizes.

However, the above-mentioned self-regulation is less significant for the late stage of isobaric model and isochoric model as shown in Fig 7(b). As the pressure jump drives a rapid shock outward at the very beginning, α-particles with ranges exceeding the hot spot radius deposit energy at the shock front instead of in the cold fuel, thus eliminate the negative feedback loop. Additionally, more energy $\Delta e$ is deposited into the shock front leads to an increased burning speed. The higher burning speeds supports the achievement of detonation. This acceleration is evident in Fig. 6(b), where the burning waves quickly develops into a detonation and the burning waves eventually catches up with the shock front.

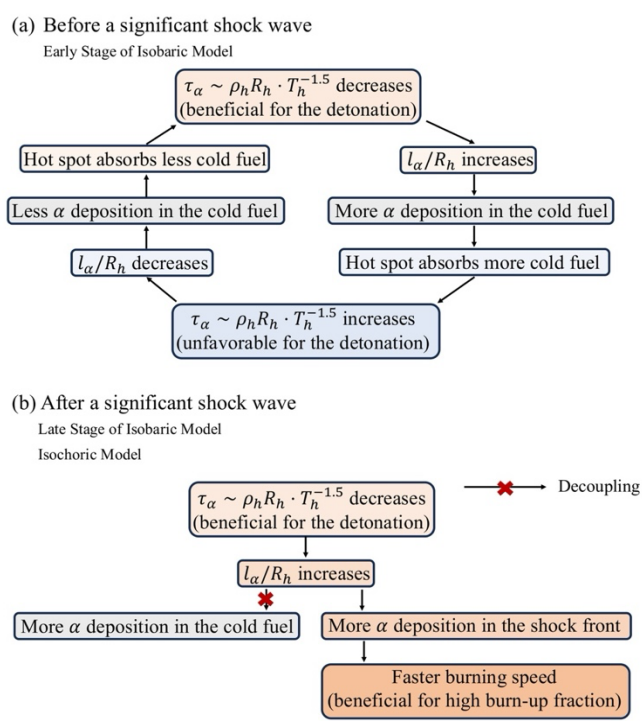

FIG 7. Schematic diagram of the self-regulation behavior at different stage for the isobaric model and isochoric model

Besides the self-regulation behavior discussed above, there is another behavior stabilizing the detonation propagation in the detonation regime. It can be illustrated as the self-modulation of density peak. Our results imply that α-particles could penetrate the high-density shock front to directly heat the unshocked cold fuel. During the heating process, an increased deposition of α-particles at the shock front leads to a higher energy input $\Delta e$. When $\Delta e$ is large, the peak density of the shock front decreases, as indicated by Eq. (13), which allows more α-particles to penetrate the peak and results in fewer α-particles deposition. This mechanism gradually stabilizes the detonation propagation. One manifestation of this mechanism is the competition between the two waves.

From above discussion, the rapidly formed shock front in the isochoric model avoids the conventional self-regulation behavior, which hinder the achievement of detonation. In the contrast, the different regulation behavior accelerates the burning wave, implying a higher burn-up fraction for the isochoric model. So, we conducted a series of simulations of the ignition process by varying deposited fast electron energies to further investigate the impact of detonation on burn-up fractions for the isochoric model. The radius of the hot spot was set to 30 μm, while the radius of the cold fuel was set to 80 μm. In Fig. 8(a), the burn-up fraction rises rapidly around 7 kJ. Meanwhile, Fig. 8(b) shows that the hot spot temperature slows down as it approaches the detonation threshold of 13.4 keV, requiring some time to surpass this value. Once the temperature exceeds the threshold, detonation occurs, and the temperature increases rapidly. However, when the time required to achieve detonation is longer, more fuel escapes, resulting in a lower burn-up fraction. From Fig. 7(b), it can be observed that the time needed to achieve detonation decreases with higher initial energy input. When the initial energy reaches 8.5 kJ, the temperature exceeds 13.4 keV quickly. As a result, for higher energy inputs, the saved time for achieving detonation is limited, so the burn-up fraction increases more gradually. For this formation of hot spot, the deposited fast electron energy of 8.5 kJ is an optimized value.

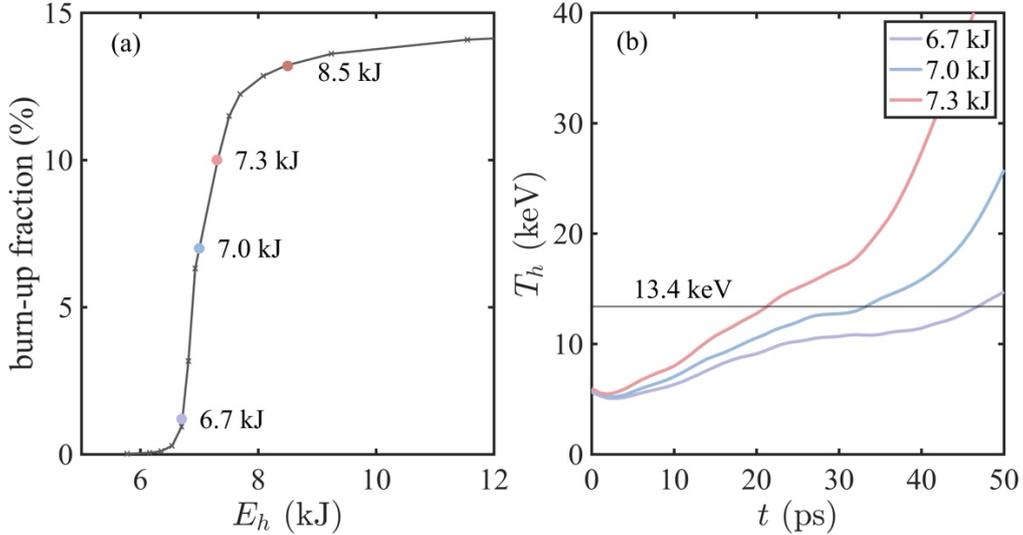

FIG 8. (a) Simulation results of the burn-up fraction with different fast electron energy input.

(b) The average temperature of hot spot for fast electron energy input of 6.7 kJ, 7.0 kJ, 7.3 kJ.

## 6. CONCLUSION

In this research, we derive a theoretical model to analyze the burning propagation process for both isobaric and isochoric model. Due to a higher density, the isochoric model usually exhibits quicker temperature rise than the isobaric model. A lower threshold temperature for detonation is given considering the α-deposition in the shock front. The new thresholds for the isochoric and the isobaric models are 13.4 keV (with a deposited fast electron energy about 8.5 kJ) and 25.1 keV, respectively. For the isochoric model, the threshold decreased by 55% compared to the previous 30 keV prediction. Notably, when detonation happens, fusion reaction immediately occurs in the newly absorbed fuel, greatly accelerating the burning process. For the isochoric model, the burning speed at the detonation state increases by 20%. The faster temperature increases and lower detonation threshold both indicates a higher burn-up fraction for the isochoric model. We conducted a series of simulations using finite fuel and identified an optimized fast electron energy input. The results confirm that the detonation would lead to a higher burn-up fraction, especially for the isochoric model, which is beneficial for fusion energy. This work also shows the possibility of studying astrophysical detonation in laboratory by isochoric fast ignition, since the new threshold is much easier to achieve.

**Acknowledgements**
This work was supported by the Strategic Priority Research Program of Chinese Academy of Sciences (Nos. XDA25010100 and XDA25051200), the National Natural Science Foundation of China (Grant Nos. 12205185), Shanghai Municipal Science and Technology Key Project (No. 22JC1401500), and Shanghai Pujiang Program (No. 22PJ1407900).